# Low temperature STM on InAs (110) accumulation surfaces

L. Canali, J.W.G. Wildöer, O. Kerkhof, and L.P. Kouwenhoven
*Department of Applied Physics and DIMES, Delft University of Technology*
*Lorentzweg 1, 2628 CJ Delft, The Netherlands*

The properties of InAs (110) surfaces have been investigated by means of low-temperature scanning tunneling microscopy and spectroscopy. A technique for *ex-situ* sulphur passivation has been developed to form an accumulation layer on such a surface. Tunneling spectroscopy at 4.2 K shows the presence of 2D subbands in the accumulation layer. Measurements in high-magnetic field demonstrate Landau quantization of the energy spectrum, both in the 2D subbands and the 3D bulk conduction band.
PACS: 73.20.Dx, 61.16.Ch, 71.70.Di

The properties of 2D electron gases (2DEG) subject to high magnetic fields have generated a large amount of theoretical and experimental work. In particular, the study of edge channels and electron-electron interactions [1] is currently a field of great interest. Spatially resolved imaging of edge channels was performed by van Haren *et al.* [2], using a special technique to obtain macroscopically wide channels. For higher spatial resolution measurements one would like to exploit the potential of STM techniques. However, this requires a semiconductor system having a 2DEG at the surface. InAs is a good candidate material for this, since it is known that it can have a surface 2DEG [3]. Wildöer *et al.* [4] have performed low-temperature scanning tunneling microscopy and spectroscopy on InAs (110) surfaces that were cleaved *in situ* at 4.2 K. A clean surface was obtained and Landau quantization was observed. In this study, however, no clear evidence was found for the formation of a surface 2DEG.

In this article we report on the development of a suitable system for STM studies on a 2DEG: the sulphur-passivated InAs (110) surface. We demonstrate the presence of 2D subbands and of Landau quantization in high magnetic fields.

From the work of Tsui [3], it is known that an oxide layer at the surface of n-type InAs pins the Fermi energy above the bottom of the conduction band, thereby forming a surface accumulation layer. We have found that oxidation of the InAs surface is not a practical ex-situ technique for an STM study, since the oxide grows too thick after just a few minutes of exposure to air. We have therefore studied different passivation techniques for InAs taking advantage of its chemical similarity with GaAs, for which a vast literature of passivation methods is available. Those techniques mainly use reactive S-containing solutions and/or gases. In practice, we have focused on two among the most practical and relatively less toxic wet techniques: one employing a solution of $(NH_4)_2S$ as in Ref. 5, the other one using a solution $CH_3CSNH_2/NH_4OH$ at 90 $^o$C as in Ref. 6.

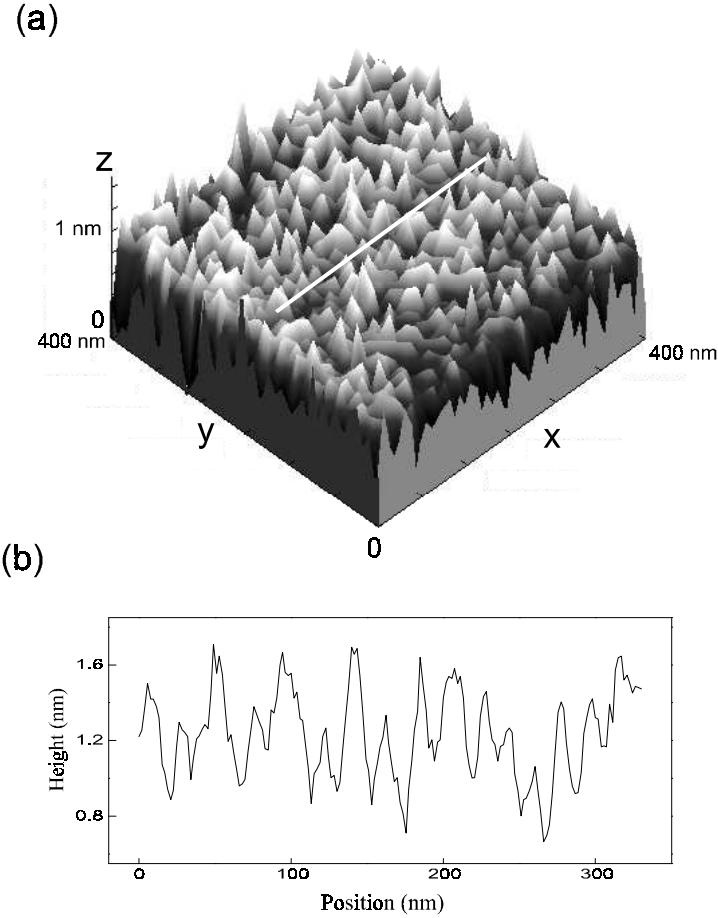

Fig.1. (a) STM topographic image at 4.2 K of the InAs (110) surface S-passivated with $CH_3CSNH_2/NH_4OH$ for 40 seconds. The image shows a rms roughness of 0.5 nm over an area of 400x400 $nm^2$. (b) Line profile along the line indicated in the topography.

We use InAs wafers with a doping concentration n = $2.4 \cdot 10^{16}$ $cm^{-3}$, corresponding to a Fermi energy $E_f$ = 13 meV above the conduction band. Passivated samples are prepared by cleaving in solution after scribing a mark on the (100) surface. We study different samples with passivation times between 5 seconds and 20 minutes. After passivation, samples are rinsed in de-ionized water and blown dry with nitrogen gas. They are then quickly mounted in our STM [7], being thus exposed to air for about 10 minutes. The STM vacuum can is then pumped out for about 90 minutes (at room temperature) in order to sublimate the excess sulphur [8]. Helium exchange gas is then allowed in the STM vacuum can just before cooling down to 4.2 K in a cryostat equipped with a 7 T superconducting magnet. The STM tips are cut in ambient from Pt(90%)Ir(10%) wire.

Figure 1 shows a typical topographic image of the S-passivated InAs surface at 4.2 K. The surface has a root-mean-square (rms) roughness of about 0.5 nm over an area of 400x400 nm$^2$. Topographies for areas of 10x10 nm$^2$ can show an rms roughness of about 0.1 nm. Despite this flatness, topographic images do not show atomic resolution. Similar results were found on GaAs by Gwo *et al.* [5]. We find that sulphur provides a near monolayer passivation, unlike the case of oxygen passivation.

Different measurements for passivation times between 5 seconds and 10 minutes show that the flattest surfaces are obtained for passivation times shorter than 2 minutes. For longer passivation times the rms roughness increases up to a few nanometers. Concerning the rms roughness, we find similar results for both of the passivation methods. From Auger spectroscopy measurements we find that the content of oxygen at the surface is significantly lower for the samples passivated with $CH_3CSNH_2/NH_4OH$ than for the ones passivated with $(NH_4)_2S$.

Figure 2 shows the I-V characteristic of a S-passivated InAs sample. The onset of the conduction band is found at about -15 meV. This is about the value of the Fermi energy in the system, as expected. The onset of the valence band is found at about -410 meV (below the Fermi energy of the system), that is in good agreement with the value of the gap of bulk InAs at 4.2 K: 420 meV. Other authors measuring STM spectroscopy on InAs cleaved *in situ* [4,9], have encountered tip-induced band bending. The actual tip-sample voltage was found to be less than the applied bias of about 0.2 V, for low-doped samples. We find that our samples do not present tip-induced band bending.

In Fig. 3 (a) a typical spectroscopy measurement at 4.2 K is reported. Three clear peaks are seen for negative bias positions: $V_0 = -131$ mV, $V_1 = -85$ mV, $V_2 = -37$ mV, and a fourth smaller peak $V_3 = -17$ mV. We ascribe these peaks to tunneling from 2D electronic subbands in the accumulation layer at the InAs surface (see Fig. 4). The exact calculation of the confinement potential in the accumulation layer requires numerical self-consistent calculations [10]. We can use a simple expression for the energy

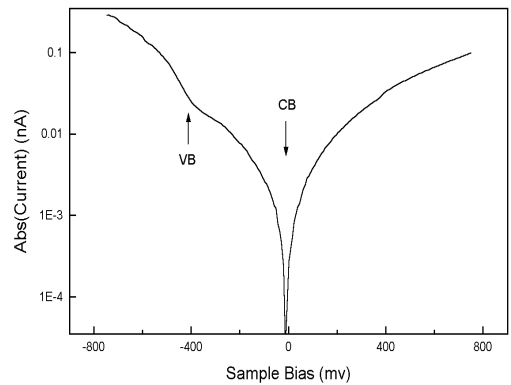

Fig.2. STM spectroscopy at 4.2 K of the InAs (110) accumulation surface S-passivated with $CH_3CSNH_2/NH_4OH$ for 40 seconds. From the I-V (and dI/dV / I/V) characteristic, the onset of the conduction and valence band are found at about -15 mV and -410 mV, respectively. This is in good agreement with the values for the Fermi energy and the gap of our InAs samples.

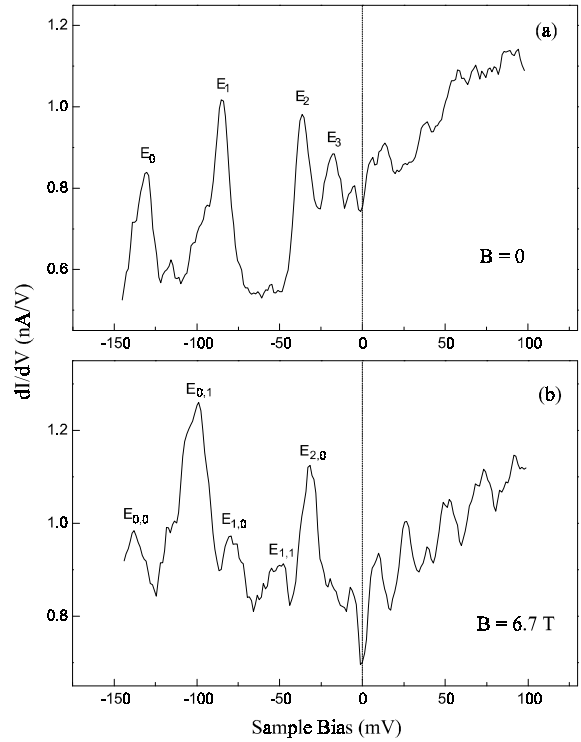

Fig.3. (a) STM spectroscopy at 4.2 K of the InAs (110) accumulation surface S-passivated with $CH_3CSNH_2/NH_4OH$ for 40 seconds. Four peaks can be resolved for negative sample bias at the values: -131, -85, -37 and -17 mV. This graph is calculated by numerical differentiation of the measured I-V characteristic.. (b) STM spectroscopy as in (a) but with a magnetic field B = 6.7 T perpendicular to the surface. Peaks at negative sample bias values of: -138, -99, -79, -49 and -32 mV are observed indicating Landau splitting of the 2D subbands. Additional peaks for positive bias values: 10, 26, 50, 73 and 91 mV are due to Landau quantization in the conduction band of InAs. The tunnel resistance is about 1 G$\Omega$. I-Vs are measured after turning the feedback control off. We average I-Vs over 25 curves each consisting of 201 points with acquisition time of 1.3 ms per point.

values of the 2D subbands when we approximate the confining potential by a triangular well [11]:

$$E_k = \left(\frac{\hbar^2}{2m^*}\right)^{1/3} \left(\frac{3\pi eF}{2}\left(k+\frac{3}{4}\right)\right)^{2/3}. \qquad (1)$$

Here F is the electric field, k = 0, 1,… is the subband index, and the zero for the energy is taken at the bottom of the well. From Eq. 1 we find $E_k / E_0 = (4k/3+1)^{2/3}$. The energy of the quantum states measured from the bottom of the conduction band is $E_k = \mathcal{E}_p + eV_k$, where $V_k$ are the measured peaks positions, and $\mathcal{E}_p$ is the Fermi-energy pinning (see Fig. 4). From the four observed peaks we obtain: $\mathcal{E}_p$ = 194±5 meV. From Eq. 1 and the calculated value for $\mathcal{E}_p$ we find for the accumulation electric field F ≈ 4·$10^6$ V/m. The accumulation length $L_a = \mathcal{E}_p / eF$ is about 50 nm. Additionally, from the measured width of the peaks we find that the broadening of the 2D subbands is about 15 meV. These values are consistent with previous work on InAs [12,13].

Peaks in the dI/dV vs. V characteristics for negative bias are observed in all our S-passivated samples. Peaks are absent in those areas where a tip crash has removed the surface layer. We find that the Fermi-energy pinning varies about 10% for different positions on the sample on distances of a few nanometers. Comparing dI/dV vs. V spectroscopy for the two passivation methods we find that samples passivated with $CH_3CSNH_2/NH_4OH$, for times less than 2 minutes, give spectroscopy characteristics with the best reproducibility and the highest resolution. In particular, on the sample that was passivated with $CH_3CSNH_2/NH_4OH$ for 40 seconds we found an area of about 10x10 nm$^2$ where the value of the Fermi-energy pinning reproduced within a few percent. Spectroscopy characteristics in that area are shown in Fig. 3.

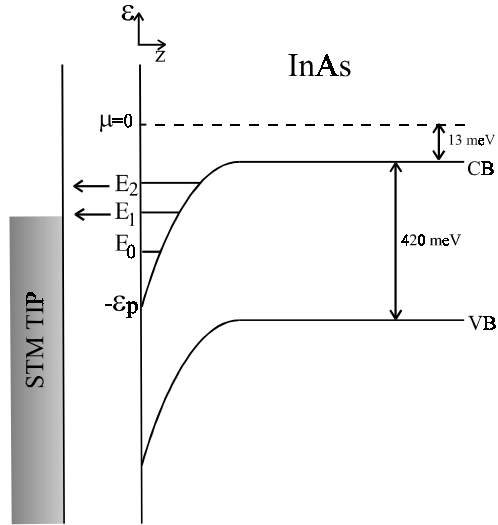

Fig.4. Schematic band diagram of n-type InAs for the case of Fermi energy pinning. The conduction band bottom at the surface is bent at the energy -$\varepsilon_p$ below the Fermi level. If the electric field in the accumulation layer is sufficiently strong, quantized 2D states are present at the surface (indicated with $E_k$). When the Fermi level of the tip is aligned with a 2D states, a peak in the dI/dV is expected.

Figure 3 (b) shows the dI/dV vs. V characteristic with a magnetic field B = 6.7 T applied perpendicular to the surface. We now discuss a simple model to interpret the oscillations present in high magnetic field characteristics. Landau quantization is expected for sufficiently strong magnetic field $\hbar\omega_c > \Gamma$, where $\hbar\omega_c = eB/m^*$ is the cyclotron energy, i.e. the energy separation between different Landau levels, and $\Gamma$ is the broadening of the Landau levels due to disorder. Landau levels have energies $E_{k,n} = E_{k,0} + \hbar\omega_c(n+1/2)$, where k = 0, 1,... is the electric field quantization number as in Eq. 1, and n = 0, 1,… is the Landau level index. The magnetic field changes also the value of the Fermi energy with respect to the bottom of the conduction band in the bulk InAs. Most probably, also the value of the electric field in the accumulation layer changes such that the bias position of the "electric peaks" ($E_{k,n=0}$) is expected to be shifted a few meV from the B = 0 peaks. We expect peaks in the dI/dV characteristic from the Landau levels at a distance of $\hbar\omega_c$(6.7 T) = 33 meV from the "electric peaks".

Comparing the two characteristics in Fig. 3 (a) and (b), respectively for B = 0 and 6.7 T, we see that the "electric peaks" present in (a) reproduce also in (b) with a shift of ±7 meV. Two extra peaks are present in (b) that can be interpreted as due to tunneling from the levels $E_{0,1}$ and $E_{1,1}$ with a measured cyclotron energy of 39 meV and 30 meV, respectively. We note that for lower magnetic field and/or for other samples the spectra manifested magnetic field dependence but no clear peak splitting was found. From this fact we find an upper boundary for the Landau level broadening $\Gamma \approx 35$ meV.

In Fig. 3 (b) peaks in the dI/dV vs. V characteristic are also present for positive sample bias. We interpret these peaks as due to Landau quantization in the conduction band of InAs. The energy spectrum of electrons in the conduction band in presence of a magnetic field directed along the z axis is $E_{k_z, n} = \hbar^2 k_z^2/2m + \hbar\omega_c(n+1/2)$, where $k_z$ is the wave vector in the z direction. The spectrum is continuous, but the density of states contains singularities at the energies $\hbar\omega_c(n+1/2)$. These singularities are also expected to give rise to peaks in dI/dV characteristics. The Fermi energy in bulk InAs, for the given electron density and B = 6.7 T, is located at about 4 meV above the first Landau level. Peaks in the dI/dV, for positive bias, are then expected to be due to tunneling of electrons into the empty Landau levels with numbers n = 1, 2,...

Figure 5 shows the dI/dV characteristic in a magnetic field B = 6.7 T for a sample passivated with $CH_3CSNH_2/NH_4OH$ for 30 seconds. Six peaks can be clearly distinguished, due to tunneling into the Landau levels n = 1 to 6. The level spacing gives a measure of the cyclotron energy $\hbar\omega_c = 23 \pm 4$ meV. From this value we can calculate the effective mass: $m^*/m_0 = 0.033 \pm 0.006$, which is higher than the value of the mass at the bottom of the conduction band: $m^*/m_0 = 0.024$. We attribute this discrepancy to mass non-parabolicity in the conduction band of InAs [12,13]. Peaks due to Landau quantization in the conduction band have been observed for different samples in magnetic fields above 6 T.

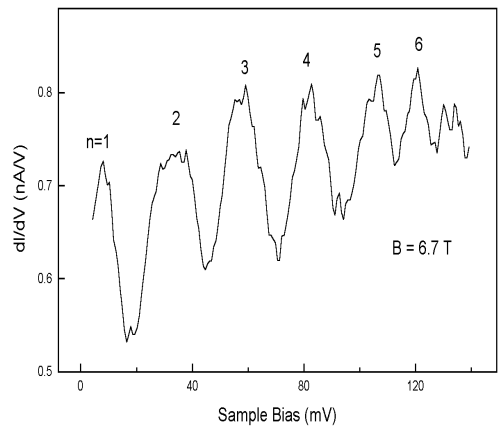

Fig.5. STM spectroscopy at 4.2 K of InAs (110) accumulation surface S-passivated with $CH_3CSNH_2/NH_4OH$ for 30 seconds. Six peaks can be resolved for positive sample bias: 8, 35, 59, 83, 106, 121 mV, indicated with n = 1,..6. They are due to tunneling of electrons into empty Landau levels in the conduction band of InAs.

In conclusion, we have developed a technique for *ex-situ* near-monolayer passivation of InAs, we have demonstrated using STM spectroscopy the presence of 2D subbands at the passivated (110) surface of InAs, and shown the presence of Landau quantization in high magnetic field. These are the first STM measurements on a 2D electron gas at the surface of a semiconductor. We have shown that our system avoids the problems of tip-induced band bending, and we have studied Landau quantization in the conduction band of InAs, where up to six Landau levels have been clearly resolved.

We gratefully acknowledge H. W. M. Salemink, J. E. Mooij, H. van Kempen, C. J. P. M. Harmans, L. J. Geerligs, and L. Gurevich for useful discussions. This research was funded by the "Stichting voor Fundamenteel Onderzoek der Materie" (FOM). L.P.K. was supported by the Royal Dutch Academy of Arts and Sciences (KNAW).

**References**
1. D. B. Chklovskii, B. I. Shklovkii, and L. I. Glazman, Phys. Rev. B **46**, 4026 (1992); A. A. Koulakov, M. M. Fogler, and B. I. Shklovskii, Phys. Rev. Lett. **76**, 499 (1996).
2. R. J. F. van Haren, W. de Lange, F. A. P. Blom, and J. H. Wolter, Phys. Rev. B **52**, 5760 (1995).
3. D. C. Tsui, Phys. Rev. Lett. **24**, 303 (1970).
4. J. W. G. Wildöer, C. J. P. M. Harmans, H. van Kempen, Phys. Rev. B **55**, R16013 (1997).
5. S. Gwo, K-J Chao, C. K. Shih, K. Sadra, and B. G. Streetman, Phys. Rev. Lett. **71**, 1883 (1993).
6. E. D. Lu, F. P. Zhang, S. H. Xu, X. J. Yu, P. S. Xu, Z. F. Han, F. Q. Xu, and X. Y. Zhang, Appl. Phys. Lett. **69**, 2282 (1996).
7. J. W. G. Wildöer, A. J. A. van Roy, H. van Kempen, and C. J. P. M. Harmans, Rev. Sci. Instrum. **65**, 2849 (1994).
8. Y. Nannichi, J. -F. Fan, H. Oigawa, and A. Koma, Jpn. J. Appl. Phys. **27**, L2367 (1988).
9. R. M. Feenstra, Phys. Rev. B **50**, 4561 (1994).
10. An-zhen Zhang, J. Slinkman, and R. E. Doezema, Phys. Rev. B **44**, 10752 (1991).
11. *Quantum semiconductor structures*, pp 19-20, C. Weisbuch, and B. Vinter, Academic press (1991).
12. D. C. Tsui, Phys. Rev. B **4**, 4438 (1971).
13. H. Reisinger, H. Schaber, and R. E. Doezema, Phys. Rev. B **24**, 5960 (1981).